\documentclass[twocolumn,prl,aps,showpacs,superscriptaddress,amsmath,amssymb]{revtex4}

\usepackage{graphicx}
\usepackage{dcolumn}
\usepackage{bm}

\newcommand{\bfr}{{\mathbf{r}}}
\newcommand{\bfQ}{{\mathbf{Q}}}
\newcommand{\bfR}{{\mathbf{R}}}
\newcommand{\bfzero}{{\mathbf{0}}}
\newcommand{\ee}{{\mathrm{e}}}

\newcommand{\dd}{{\mathrm{d}}}
\newcommand{\Hkin}{{H_{\mathrm{kin}}}}
\newcommand{\barR}{\overline{\bfQ}}

\begin{document}

\preprint{APS/123-QED}

\title{Effect of vibrations on the pre-edge features 
  of x-ray absorption spectra}

\author{Christian Brouder, Delphine Cabaret,
Am\'elie Juhin and Philippe Sainctavit}
\affiliation{%
Institut de Min\'eralogie et de Physique des Milieux Condens\'es,
CNRS UMR 7590, Universit\'es Paris 6 et 7, IPGP, 140 rue de Lourmel,
75015 Paris, France.
}%

\date{\today}

\begin{abstract}
The influence of atomic vibrations on x-ray absorption 
near edge structure (XANES) is calculated by assuming
that vibrational energies are small with respect to the
instrumental resolution. 
The resulting expression shows that, at the $K$-edge, vibrations
enable electric dipole transitions to $3s$
and $3d$ final states. 
The theory is applied to the $K$-edge of Al in $\alpha$-Al$_2$O$_3$
and of Ti in TiO$_2$ rutile and compared with experiment.
At the Al $K$-edge, sizeable transitions towards $3s$ final
states are obtained, leading to a clear improvement of the agreement
with experimental spectra.
At the Ti $K$-edge, electric dipole transitions towards $3d$ final
states explain the temperature dependence of the
pre-edge features.
\end{abstract}

\maketitle

Vibronic coupling describes the interaction between
electrons and atomic motions.
It plays a prominent role in optical spectroscopy where
it is the source of the color of many pigments and 
gemstones~\cite{Bridgeman-97-Intens}. For instance,
the red color of Black Prince's ruby is due to ``d-d" transitions
of chromium impurities in a spinel crystal. But these transitions are
forbidden because chromium occupies an inversion center
of the spinel lattice. They become allowed when vibrations
break inversion symmetry.

In the x-ray range, vibrations far from the edge are taken into
account through a Debye-Waller factor
$\ee^{-2 k^2\sigma^2}$~\cite{Vila-07}.
If the validity of this factor is assumed to extend 
to the near-edge region, where $k\simeq0$, then
vibrations seem negligible in XANES spectra.

However, three arguments indicate that vibronic coupling
can be sizeable in the XANES region:
(i) Vibronic coupling was detected by x-ray resonant
  scattering experiments at the Ge $K$-edge~\cite{Kirfel,Dmitrienko-05};
(ii) Some XANES peaks seem to be due to forbidden transitions
  to $3s$ states, a prominent example being the Al $K$-edge 
  in minerals~\cite{Li-95};
(iii) A temperature dependence of the pre-edge structure was
  observed at the Ti $K$-edge in SrTiO$_3$~\cite{Nozawa}
  and TiO$_2$~\cite{Durmeyer-10}.

In the optical range, the effect of vibrations is usually
taken into account through the Franck-Condon factors. 
In the x-ray range, Fujikawa and
coll. showed in a series of papers of increasing 
sophistication~\cite{Fujikawa-96-JPSJ,Fujikawa-99,Arai-XAFS13-phonon}
that the effect of the Franck-Condon factors can be
represented by the convolution of the 
``phonon-less'' x-ray absorption spectrum with
the phonon spectral function. Such a convolution
leads to a broadening of the peaks with increasing
temperature but this effect is hardly observable
in the pre-edge region.

Moreover, in the x-ray range
it was shown that the large core-electron-phonon coupling 
of the $1s$ core hole of carbon in diamond induces a 
strong lattice distortion
and significant anharmonic contributions~\cite{Mader}.

The key observation is that all these effects can be
easily taken into account if the vibrational energies are
small with respect to the XANES spectral resolution
(core hole lifetime + instrumental resolution). This condition 
is certainly not satisfied 
at the C $K$-edge~\cite{Mader} but it becomes reasonable
at the Al and Ti $K$-edges.
In that range the XANES resolution is around one eV
whereas the energy of vibrational
modes is of the order of a few hundredths of eV
although, of course, several phonons can be simultaneously
present.

In this paper, we first use this observation
to derive a manageable expression for the 
vibronically-coupled x-ray absorption spectra.
Then, we apply the so-called \emph{crude Born-Oppenheimer}
approximation to further
simplify this expression, so that only the core-hole
motion remains. The resulting equation is compared
with experiment in two different cases.
At the Al $K$-edge in $\alpha$-Al$_2$O$_3$ (corundum), vibrations
induce transitions to $3s$ final states.
These ($1s\rightarrow 3s$) monopole transitions
explain a pre-edge peak that is completely absent from standard calculations.
At the Ti $K$-edge in TiO$_2$,  $1s\rightarrow 3d$ transitions
are induced by vibrations. This explains why only the
first two pre-edge peaks grow with temperature.

\emph{XANES formula within the Born-Oppenheimer framework.}
According to the Born-Oppenheimer 
approximation~\cite{Born-1927,Henderson}, the
wavefunction of the system of electrons and nuclei can
be written as the product
$\chi_n^j(\barR)\psi_n(\bfr;\barR)$, where 
$\bfr$ is the electronic variable 
and $\barR=(\bfQ_1,\dots,\bfQ_N)$ collectively denotes the 
position vectors of the $N$ nuclei of the system.
The electronic wavefunction $\psi_n(\bfr;\barR)$,
with energy $\epsilon_n(\barR)$,
describes the state of the electrons
in a potential where the nuclei are fixed at position
$\barR$. The ground state corresponds to $n=0$.
The origin of the nuclear variables is
chosen so that $\barR=\bfzero$ is the
equilibrium position, i.e. $\epsilon_0(\bfzero)$
is the minimum of $\epsilon_0(\barR)$.

For each $n$, the vibrational wavefunctions $\chi_n^j(\barR)$
are the orthonormal solutions of the Schr{\"o}dinger equation
\begin{eqnarray*}
\big(\Hkin(\barR)+\epsilon_n(\barR)\big)\chi^j_n(\barR)
 &=& E_n^j \chi^j_n(\barR).
\end{eqnarray*}
The total energy of the electrons + nuclei system is
$E_n^j$.
Within the Born-Oppenheimer approximation, the x-ray
absorption cross-section is 
\begin{eqnarray*}
\sigma(\omega) &=& 4\pi^2 \alpha_0 \hbar\omega
\sum_{fj} \Big| \int \dd \barR \dd \bfr
  \chi^j_f(\barR)^*\psi_f(\bfr,\barR)^* \varepsilon\cdot\bfr
\\&&
  \times \chi_0(\barR)\psi_0(\bfr,\barR)\Big|^2
  \delta(E_f^j-E_0-\hbar\omega),
\end{eqnarray*}
where $\alpha_0$ is the fine structure constant,
$\hbar\omega$ the energy of the incident x-rays
and $\varepsilon$ their polarization vector.
The core-hole lifetime and the instrumental resolution
can be represented by the convolution of the absorption
cross-section with a Lorentzian function
$(\Gamma/\pi)/(\omega^2+\Gamma^2)$.
This gives us
\begin{eqnarray*}
\sigma_\gamma(\omega) &=& 4\pi \alpha_0 
\sum_{fj} \Big| \int \dd \barR \dd \bfr
  \chi^j_f(\barR)^*\psi_f(\bfr,\barR)^* \varepsilon\cdot\bfr
\\&&
 \times \chi_0(\barR)\psi_0(\bfr,\barR)\Big|^2
  \frac{(E_f^j-E_0^0)\gamma}{(E_f^j-E_0-\hbar\omega)^2 + \gamma^2},
\end{eqnarray*}
where $\gamma=\hbar\Gamma$.

The energy $E_f^j$ can be written as the
sum of the electronic energy at equilibrium position
$\epsilon_f$ and a vibrational energy
$E_f^j=\epsilon_f+ E^{fj}_{\mathrm{vib}}$.
When $\gamma$ is much larger than 
the vibrational energy 
we can neglect the contribution of
$E^{fj}_{\mathrm{vib}}$ and sum over the vibrational
states $\chi^j_f(\barR)$.
The completeness relation gives us
\begin{eqnarray*}
\sum_j \chi^j_f(\barR)^* \chi^j_f(\barR')
&=& \delta(\barR-\barR').
\end{eqnarray*}
Therefore,
\begin{eqnarray}
\sigma_\gamma(\omega) &=& 4\pi \alpha_0 
\int \dd \barR 
\sum_f 
\Big|
\int \dd \bfr
  \psi_f(\bfr,\barR)^* \varepsilon\cdot\bfr
\nonumber\\&&
  \times \chi_0(\barR)\psi_0(\bfr,\barR)\Big|^2
  \frac{(\epsilon_f-\epsilon_0)\gamma}
  {(\epsilon_f-\epsilon_0-\hbar\omega)^2 + \gamma^2}.
\label{sigmagamma}
\end{eqnarray}
Note that we derived this result without making the harmonic
approximation. Therefore, the possible anharmonic behavior
due to the core hole~\cite{Mader} is taken into account.
Note also that the final and initial energies 
$\epsilon_f$ and $\epsilon_0$ do not depend on $\barR$.
Therefore, eq.~(\ref{sigmagamma}) is not the average 
of standard XANES spectra over various nuclear positions
$\barR$. In other words, we have here a way to distinguish
thermal disorder from static disorder due to impurities
and structural defects.

Now we make a different approximation for the initial
and final electronic states. 
For a $K$-edge, the $1s$ core level wavefunction is highly 
localized around the nucleus and it weakly depends on
the surrounding atoms. Therefore, we can approximate
$\psi_0(\bfr,\barR)$ by $\phi_0(\bfr-\bfQ_a)$, where
$\bfQ_a$ is the position vector of the absorbing atom
and where $\phi_0$ is the $1s$ wavefunction of the absorbing atom 
at equilibrium position.
For the final electronic states $\psi_f(\bfr,\barR)$
in the presence of a core hole, the
variation of the nuclear coordinates $\barR$ in eq.~(\ref{sigmagamma})
is ruled by the vibrational wavefunction $\chi_0$ of the initial state,
which is expected to be rather smooth.
Therefore, we make the standard
\emph{crude} Born-Oppenheimer approximation, according
to which the electronic wavefunction does not significantly
vary with $\barR$ for small vibrational motions.
In other words,
$\psi_f(\bfr,\barR)\simeq \phi_f(\bfr)$,
where $\phi_f(\bfr)=\psi_f(\bfr,\bfzero)$.
This gives us
\begin{eqnarray*}
\sigma_\gamma(\omega) &=& 4\pi \alpha_0 \hbar\omega 
\int \dd \barR |\chi_0(\barR)|^2
\sum_f 
\Big|
\int \dd \bfr
  \psi_f(\bfr)^* \varepsilon\cdot\bfr
\\&&
  \times \phi_0(\bfr-\bfQ_a)\Big|^2
  \frac{\hbar\gamma}
  {(\epsilon_f-\epsilon_0-\hbar\omega)^2 + \gamma^2}.
\end{eqnarray*}
When the crude Born-Oppenheimer approximation is not valid, 
it is possible
to Taylor-expand $\psi_f(\bfr,\barR)$ as a function of 
$\barR$~\cite{BornHuang}.
The integral over electronic variables depends only
on the position of the absorbing atom, from now on
denoted by $\bfR$. Therefore,
we can integrate over the other nuclear variables 
and the expression becomes
\begin{eqnarray}
\sigma_\gamma(\omega) &=& 4\pi \alpha_0 \hbar\omega 
\int \dd \bfR \rho(\bfR)
\sum_f 
\Big|
\int \dd \bfr
  \psi_f(\bfr)^* \varepsilon\cdot\bfr
\nonumber\\&&
  \times \phi_0(\bfr-\bfR)\Big|^2
  \frac{(\epsilon_f-\epsilon_0)\gamma}
  {(\epsilon_f-\epsilon_0-\hbar\omega)^2 + \gamma^2},
\label{intR}
\end{eqnarray}
where
$\rho(\bfR)=
\int \dd \barR |\chi_0(\barR)|^2 \delta(\bfQ_a-\bfR)$.
Within the harmonic approximation,
the core displacement distribution has 
the form~\cite{Maradudin}.
\begin{eqnarray*}
\rho(\bfR) &=& \exp\Big( - \frac{\bfR \cdot U^{-1} \cdot\bfR}{2}\Big),
\end{eqnarray*}
where $U$ is the thermal parameter matrix~\cite{Giacovazzo}
that is measured in x-ray or neutron scattering
experiments.

\emph{Calculation of the matrix element}.
For a hydrogenoid atom, the $1s$ core-hole radial
wavefunction is proportional to $\ee^{-a r}$,
where $a = Z/a_0$, $Z$ is the
atomic number and $a_0$ the Bohr radius.
The $1s$ wavefunction $\phi_0(r)$ of a true atom is close to 
that of a hydrogenoid one and can be written as
a fast converging linear combination of exponentials.
Thus, $\phi_0(\bfr-\bfR)$ becomes a sum of
shifted exponentials that can be described by 
the Barnett-Coulson expansion \cite{Barnett}
\begin{eqnarray*}
\ee^{-a|\bfr-\bfR|}
&=&
  \sum_{n} (2n+1) P_n(\hat\bfr\cdot\hat\bfR)
   c_n(r,R),
\end{eqnarray*}
with $P_n$ a Legendre polynomial and
\begin{eqnarray*}
c_n(r,R)
&=&
  -\frac{1}{\sqrt{r R}}
  \Big(r_< I'_{n+1/2}(a r_<) K_{n+1/2}(a r_>)
\\&&
  +r_>I_{n+1/2}(a r_<) K'_{n+1/2}(a r_>)\Big),
\end{eqnarray*}
where $r_<$ ($r_>$, resp.) is the smaller (larger, resp.) of
$r$ and $R$, $I_\nu(z)$ and $K_\nu(z)$ are the modified Bessel
functions and 
$I'_\nu(z)$ and $K'_\nu(z)$ their derivatives with respect to $z$.
For notational convenience, we consider that the core wavefunction
can be represented by a single exponential $\phi_0(r)=C \ee^{-ar}$.

To calculate the matrix element, we expand the final state
wavefunction over spherical harmonics
$\psi_f(\bfr)=\sum_{\ell m} f_{\ell m}(r) Y_\ell^{m}(\hat\bfr)$.
The matrix element over the electronic variable is
\begin{eqnarray*}
\int \dd \bfr
  \psi_f(\bfr)^* \varepsilon\cdot\bfr
  \phi_0(\bfr,\bfR)
&=&
\sum_{\ell m} X_\ell^m(\bfR),
\end{eqnarray*}
with
\begin{eqnarray*}
X_\ell^m(\bfR) &=& 
C\sum_{n=0}^\infty
\int\dd\bfr
  f_{\ell m}^*(r) Y^m_\ell(\hat\bfr)^*
  \varepsilon\cdot\bfr 
 \\&& (2n+1) c_n(r,R) P_n(\hat\bfr\cdot\hat\bfR).
\end{eqnarray*}
Standard angular momentum recoupling leads to
\begin{eqnarray}
X_\ell^m(\bfR) &=& 
 \frac{(4\pi)^2 C}{3}
\sum_{n=|\ell\pm1|}
\int r^3 \dd r
  f_{\ell m}^*(r) c_n(r,R)
\nonumber\\&&
  \sum_{\lambda} 
(-1)^m Y_1^{-\lambda}(\varepsilon) Y_n^{\lambda-m}(\hat\bfR)
 C^{\ell m}_{1\lambda n m-\lambda},
\label{Xellm}
\end{eqnarray}
where 
$C^{k\kappa}_{1\lambda np}$ are Gaunt coefficients.
Equation~(\ref{Xellm}) shows that all values of
the final state angular momentum $\ell$ are now allowed.
The core-hole wavefunction is still spherical, but with
respect to a shifted centrum. Thus, with respect to the
original spectrum, it is a sum over all angular momenta
given by the Barnett-Coulson expansion.
Thus, all final states angular momenta are available in spite
of the fact that only electric dipole transitions are allowed.
In particular, vibrations allow for dipole transitions
to the $3s$ and $3d$ final states at the $K$-edge.

\emph{General features of vibrational transitions}.
The foregoing approach enables us to draw some general
conclusions concerning the effect of vibrations on
XAS pre-edge structure.
This effect is measurable if the density of non-$p$ states
of the system in the final state (i.e. in the presence
of a core hole) is large and well localized near the
Fermi energy (vibrational transitions towards $p$-states
would be masked by the allowed vibrationless transitions).
For example, just above the Fermi level, 
many aluminum or silicon compounds have
a strong density of $3s$ states and many
transition metal compounds have a large density of
$3d$ states. In the first case, vibrational transitions
appear as monopole $1s\to 3s$ transitions, which
are completely excluded with electromagnetic 
transitions. In the second case, vibrational transitions
superimpose upon electric quadrupole $1s\to 3d$ transitions.
Thus, vibrations induce
transitions at specific energies in the pre-edge but
hardly modify the rest of the XANES spectrum.

A temperature dependence of vibrational transitions
is expected if $U$ varies with temperature. 
This occurs between 0~K and room temperature
if the sample has soft modes (i.e. low energy phonons).
Otherwise, vibrational transitions are only due
to the zero-point motion of the nuclei. In other
words, vibrational transitions are then a consequence of
the fact that, even at 0~K, nuclei are not
localized at a single point.

We test these conclusions with two 
examples: the Al $K$-edge in corundum (where
vibrational transitions to $3s$ states are expected)
and the Ti $K$-edge in rutile (where 
vibrational transitions to $3d$ states and a temperature
dependence are expected).

\emph{The Al $K$-edge in corundum}.
The X-ray absorption cross section within the crude Born-Oppenheimer
approximation has been implemented in the XSpectra 
package~\cite{XSpectra}
 of the Quantum-espresso suite of codes~\cite{Giannozzi-09}.
To calculate the integral over $\bfR$ in eq.~(\ref{intR}),
it is found sufficient to compute the integral over a cube of
size 2$\Lambda$, where $\Lambda$ is the largest eigenvalue of the matrix $U$.
This cube is cut into 27 smaller cubes where the integral
is carried out using eq.~(25.4.68) of Ref.~\onlinecite{Abramowitz}. 
The technical details of the self-consistent calculation are
the same as in Refs.~\cite{Cabaret-05,Cabaret-10}.

\begin{figure}
\includegraphics[width=8.0cm]{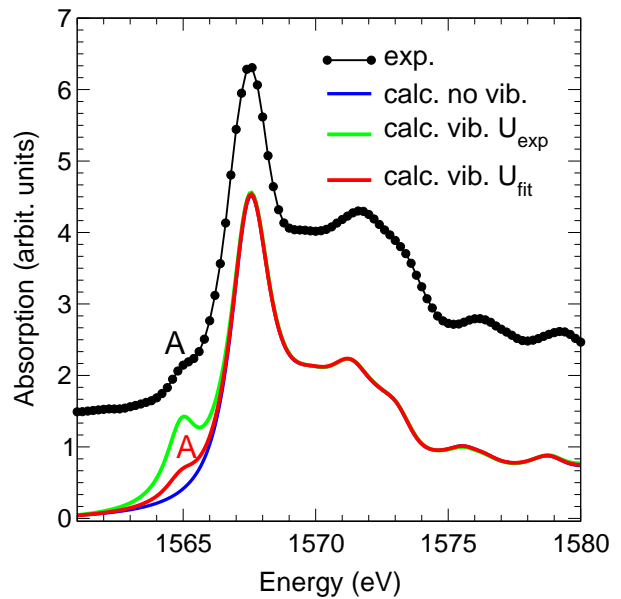}
\caption{Experimental~\cite{Ildefonse} and calculated Al $K$-edge isotropic
spectrum of corundum at 300~K. $U_{\mathrm{exp}}$ and
$U_{\mathrm{fit}}$ are the experimental and fitted 
thermal parameters (see text).
\label{figAl}}
\end{figure}
Figure~\ref{figAl} shows the result for 
experimental mean displacements $\sigma_1=\sigma_2=$0.048~\AA,
$\sigma_3 =$0.049~\AA~\cite{Maslen-93}. The mean displacements
$\sigma_i$ are the square root of the eigenvalues
of $U$ and they have a more direct physical meaning than $U$.
The vibration transitions are observed exactly
at the position of the pre-edge peak which is absent
from the calculation without vibrations.
However, the vibrational transitions are overestimated
and a better agreement is obtained when setting
the mean displacements to 0.026~\AA.
Therefore, computed vibrational transitions show up at the
right position but with a too large intensity.
A part of this discrepancy might be due to
the fact that the experimental thermal parameters include
some amount of static disorder, but most of it
probably comes from the crude Born-Oppenheimer
approximation. 
The main vibrational transitions come
from the \emph{relative} displacement of 
the aluminum atom with respect to the six oxygen first
neighbors. 
The  vibrational modes
involving an overall motion of the aluminum
atom with its oxygen octahedron contribute
to the thermal parameter $U$ although
they do not contribute
to the vibrational transitions.
Because of the crude Born-Oppenheimer approximation,
the vibrations are summarized in the thermal parameter,
which overestimates the real effect of vibrations.
A similar phenomenon occurs with the EXAFS Debye-Waller
factor which is different from the one determined by x-ray diffraction
because only relative displacements must be taken into account.
This point will now be confirmed with the
Ti $K$-edge absorption in rutile.

\emph{The Ti $K$-edge of rutile}.
With the Ti $K$-edge of rutile, we test the limit of 
the crude Born-Oppenheimer approximation.
Indeed, this approximation assumes that the final state
wavefunction does not change when the crystal vibrates.
This might be reasonable for the Al $3s$ states
because they are poorly localized and overlap the
oxygen $2p$ orbitals, but this is not
true for the $3d$ states of Ti in rutile, which
are localized near the Ti nucleus.
\begin{figure}
\includegraphics[width=8.0cm]{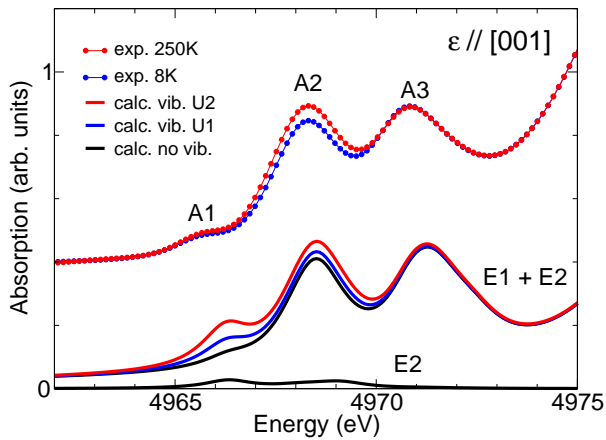}
\caption{Experimental~\cite{Durmeyer-10} and calculated Ti $K$
pre-edge spectrum of rutile. 
$E1$ and $E2$ are electric dipole and quadrupole transitions.
\label{figTipara}}
\end{figure}


Figure~\ref{figTipara}
shows the experimental and theoretical Ti $K$-edge spectrum 
of rutile with the polarization parallel
to the $c$ axis and at two temperatures.
A similar result is obtained when $\varepsilon$
is perpendicular to the $c$ axis.
The mean displacements $\sigma_1,\sigma_2,\sigma_3$ used
in the calculations are, in \AA, 0.0028, 0.0027, 0.0023 for $U1$ (8~K)
and 0.0043, 0.0042, 0.0035 for $U2$ (250~K). They have been chosen to 
approximate the intensity of peak A2
at low and room temperatures, respectively.
These values are more than ten times smaller 
than the experimental values~\cite{Burdett-87}.
Moreover, when the thermal factor is adjusted to be
consistent with the temperature variation of the second
peak, then the temperature variation of the first peak
is overestimated. This can probably be
attributed to the crude Born-Oppenheimer approximation.
Despite these drawbacks, several aspects of the 
experimental vibrational transitions are correctly reproduced:
(i) only the first and the second peaks exhibit a
temperature dependence, the rest of the spectrum is not
modified; (ii) the peaks do not shift and do not broaden;
(iii) the peaks increase with temperature. 

This work was performed using HPC resources from
GENCI grant 2009-2015 and 1202.

%

\end{document}